# 21st Century Planar Field Emission Theory and its Role in Vacuum Breakdown Science

Richard G. Forbes

University of Surrey, Advanced Technology Institute & Dept. of Electrical and Electronic Engineering, Guildford, Surrey GU2 7XH, UK

***Abstract*—For explaining electrical breakdown, field electron emission (FE) is a mechanism of interest. In the period 2006 to 2010 there were significant developments in basic FE theory, but these have not yet fully entered general thinking in technological FE areas, which are often still based on 1960s thinking or (in some contexts 1920s thinking) about FE theory. This paper outlines the history of FE theory and provides an overview of modern developments and of some related topics, in so far as these affect the interpretation of experiments and the explanation of physical phenomena. The paper concentrates on principles, with references given where details can be found. Some suggestions are made about moving to the use of "21st-Century" FE theory. In addition, an error in Feynman's treatment of the electrostatics of pointed conductors is displayed, and it is found that Zener tunneling is implausible as a primary cause of vacuum breakdown from a CuO overlayer.**

## I. General Introduction

*Fowler-Nordheim (FN) tunneling* is the process whereby electrons escape by wave-mechanical tunneling through a rounded triangular potential-energy (PE) transmission barrier. The barrier is formed by applying a high negative local surface electrostatic field, typically of magnitude a few V/nm (a few GV/m), to an "emitter" surface. *Field electron emission (FE)* is a statistical emission regime in which most electrons escape from emitter states significantly below the top of the transmission barrier. With metal emitters, conventional theory assumes that electrons near the emitter surface are in local statistical-mechanical equilibrium at a well-defined thermodynamic temperature $T$. In most cases the theory given is that for $T$= 0 K, because the room-temperature correction is very small. This zero-temperature ("zero-K") theory is of primary interest here.

Field electron emission has theoretical value as one of the paradigm examples of quantum-mechanical tunneling, but its main practical interest is as an electron emission process used or occurring in a very wide range of established and potential technological applications. Obviously, the context here is electrical breakdown in vacuum.

The FE theory used in most technological contexts is based on an underlying physical model introduced into FE by Fowler and Nordheim in 1928 [1]. This uses Sommerfeld-type free-electron theory, disregards atomic structure, and assumes that emitters can be modeled as having classically smooth planar surfaces of large lateral extent. I call this *smooth planar metal-like emitter (SPME)* methodology.

Some modern FE theories consider non-planar emitters, or the effects of putting atomic wave-functions into modeling, and thus go beyond SPME methodology. Theoretical models of this kind are useful in simulations and theoretical explorations, but at present it is not easy to use most of them for consensual interpretation of practical results. For interpretation of experimental results and extraction of characterization parameters, it is easier (at present) to use a modern ("21st Century") version of SPME methodology.

Modern vacuum-breakdown discussions mostly use versions of FE theory developed in the 1950s and 1960s. However, in the period 2006 to 2010 there were significant advances in FE theory. This paper provides an overview of these advances, in particular as they affect the interpretation of experimental phenomena. The paper concentrates on principles, but indicates where details can be found.

In FE there is limited consensus on notation systems. For consistency, the author has converted equations found in other papers (including his own earlier papers) into his currently preferred notation system.

## II. The Historical Phases of FE Theory

In SPME methodology, *FE core equations* give a formula for the *local emission current density (LECD)* $J_L$, at some location "L" on the emitter surface, in terms of the local work function $\phi$ and the magnitude $F_L$ of the local surface electrostatic field at this location (or alternatively the so-called "scaled field" discussed below). Different historical phases of FE theory are characterized by different thinking about the "most useful" form of core equation.

Leaving aside the pre-history phases where the phenomenon of "auto-electronic emission" was known but not understood, the first phase started with a FE core equation given by Fowler and Nordheim (FN) in 1928 [1].

The FN paper contains several errors. In particular, (22) in [1] contains a large numerical error, roughly around $10^{17}$, subsequently corrected [2]. A simplified version of (1) in [2], used in [3] (as part of more sophisticated arguments), is

$$J_L^{el} \equiv a\phi^{-1}F_L^2 \exp[-b\phi^{3/2}/F_L] \,, \qquad (1)$$

where $a$ [≅ 1.541434 μA eV V$^{-2}$] and $b$ [≅ 6.830890 eV$^{-3/2}$ GV m$^{-1}$] are universal constants now usually known as the first and second FN constants [4,5]. Equation (1) is sometimes called the *elementary FE equation*. Following an (erroneous) suggestion in [6] that (1) above was a good approximation, this equation has become widely used in field emitter array literature over the last 20 years.

The second FE theory phase started with a paper by Burgess, Kroemer and Houston (BKH) [7] in 1953. BKH had discovered a mathematical error in a 1928 paper by Nordheim [8]. Nordheim had discovered a conceptual error (underestimating image-PE effects) in the FN 1928 paper and was attempting to correct this (see [9] for details). This BKH finding means that the theory in ALL published field emission material prior to 1953 (including the 1933 Sommerfeld and Bethe article [10] that was a foundational document for modern condensed matter physics) is quantitatively incorrect.

This BKH result was then used by Murphy and Good (MG) in 1956 [11] to develop a revised FE core equation. The MG derivation was published using a mixture of Gaussian-system and Hartree-unit-system equations and can be difficult for younger researchers to follow: for a modern derivation see [12]. The essential difference about the MG treatment is as follows: it uses SPME methodology but assumes a Schottky-Nordheim (SN) ("planar image-rounded") transmission barrier, rather than the "exactly triangular" barrier assumed by FN.

It is convenient here to write the *Murphy-Good zero-temperature core FE equation* in the "linked" form

$$J_{\mathrm{L}}^{\mathrm{MG0}} = \mathrm{t_F}^{-2} J_{\mathrm{kL}}^{\mathrm{SN}} , \qquad (2a)$$

$$J_{\mathrm{kL}}^{\mathrm{SN}} \equiv a\phi^{-1}F_{\mathrm{L}}^{2} \exp[-\mathrm{v_F} b\phi^{3/2}/F_{\mathrm{L}}] , \qquad (2b)$$

where $J_{\mathrm{kL}}^{\mathrm{SN}}$ is defined by (2b) and is called the *kernel current density for the SN barrier*. The parameters $\mathrm{v_F}$ and $\mathrm{t_F}$ are particular values of the *field emission special mathematical functions* "v" and "t". These act as correction factors to the pre-exponential and the exponent of the elementary FE equation, and are discussed further below.

MG [11] gave a definition of $\mathrm{v_F}$ in terms of complete elliptic integrals. This is not convenient for practical use and many (around 15-20) different algebraic approximations—mostly now obsolete—have been developed. Almost all of these formulae were stated as functions of the *Nordheim parameter* $y_{\mathrm{F}}$ (for a barrier of zero-field height $\phi$) given by

$$y_{\mathrm{F}} \equiv c\,\phi^{-1} F_{\mathrm{L}}^{1/2} \equiv (e^{3}/4\pi\varepsilon_{0})^{1/2} \quad \phi^{-1} F_{\mathrm{L}}^{1/2}, \qquad (3)$$

where $c$ [$\equiv (e^{3}/4\pi\varepsilon_{0})^{1/2} \cong 3.794686\times10^{-5}$ eV V$^{-1/2}$ m$^{1/2}$] is the *Schottky constant*. For example, the 1997 SLAC Report [13] uses the Charbonnier and Martin approximation [14]

$$\mathrm{v_F} \approx \mathrm{v_{CM}}(y_{\mathrm{F}}) = 0.956 - 1.062\, y_{\mathrm{F}}^{2}. \qquad (4)$$

The third FE theory phase has started with the 2006 discovery [15], with Dr Jonathan H. B. Deane, of a "simple good approximation" for $\mathrm{v_F}$ (see Section IIIB) and realization that the *scaled field* $f$ [$=y_{\mathrm{F}}^{2}$] is more natural as a modeling variable. This phase has developed untidily over several years, and now comprises the following changed approaches – not all of which have yet been widely accepted.

## III. 21ST CENTURY SPME EMISSION THEORY

### *A. The special mathematical function* v($x$)*: Introduction*

A process is underway to recognize that the function "v" and related functions are special mathematical functions (SMFs), and to separate the pure mathematics of these functions from their use in modeling. It has been found [16] that "v" is a very special solution of the Gauss Hypergeometric Differential Equation (HDE). This is its highest-mathematical-level definition. Thus, the best mathematical approach is to express "v" as a function of the independent variable in the Gauss HDE. This variable is denoted here by "$x$", and called the *Gauss variable*. Thus, I now write this function as v($x$). In accordance with international scientific typesetting rules for SMFs [17], the symbol "v" is typeset upright.

In older literature, the SMF "v" is written as a function of the Nordheim parameter $y$ [$=+x^{1/2}$], as $v(y)$. From a mathematical viewpoint, this older approach should now be regarded as *mathematically perverse*, because it is **not** normal mathematical practice to express a SMF as a function of the *square root* of the independent variable in its defining equation.

A proof can be given (see [5], and below) that, in the MG FE equation, the function v($x$) is applied to tunneling through a SN barrier of zero-field height $\phi$ by setting $x$ equal to the *scaled-field* $f$ (for a barrier of zero-field height $\phi$) defined by

$$f = c^{2}\phi^{-2}F_{\mathrm{L}} \approx [1.439965\ \mathrm{eV}^{2}\,(\mathrm{GV/m})^{-1}]\cdot\phi^{-2}F_{\mathrm{L}} . \qquad (5a)$$

Thus, $\mathrm{v_F} = \mathrm{v}(x{=}f)$. In MG FE theory, this definition can also be written in terms of the *reference field* $F_{\mathrm{R}}$ [$=c^{-2}\phi^{2}$] needed to pull the top of the SN barrier down to the Fermi level, as

$$f = F_{\mathrm{L}}/F_{\mathrm{R}} . \qquad (5b)$$

For some purposes in modeling, for example the theory of high temperature emission [18], it may be slightly more convenient to set $x{=}y_{H}^{2}$, where $y_{H} = cH^{-1}F_{\mathrm{L}}^{1/2}$, where $H$ is the zero-field height of the barrier of interest. But for discussion of current-voltage characteristics, the scaled field $f$ is a much more useful parameter than $y_{\mathrm{F}}$.

### *B. Mathematical forms and approximations for* v($x$).

Because v($x$) is a special solution of the Gauss HDE, it follows that v($x$) has many equivalent exact mathematical definitions and expressions. FE literature now contains:

- a definition in terms of a defining equation;
- an exact analytical expression;
- two forms of exact series expansion;
- three forms of definition in terms of integrals;
- three definitions in terms of complete elliptic integrals.

Useful modern approximations include:

- a 10-term high-precision algebraic approximation, with maximum error magnitude less than $8\times10^{-10}$ over the range $0\leq x\leq1$ (see Appendix to [5]);
- a 3-term "simple good approximation" [15], with maximum error magnitude 0.0024 (0.33%) over the range $0\leq x\leq1$; this has the form

$$\mathrm{v}(x) \approx \mathrm{v_{F06}}(x) = 1 - x + (1/6)x\ln x. \qquad (6)$$

Further details can be found in [5]. Comparative details of many of the older approximations, which have mostly been made obsolete by the 21st-Century approximations noted above, can be found in [19].

The lowest terms of the simplest exact expansion are [16]

$$\mathrm{v}(x) = 1 - \left(\tfrac{9}{8}\ln 2 + \tfrac{3}{16}\right)x - \left(\tfrac{27}{256}\ln 2 - \tfrac{51}{1024}\right)x^{2} - \left(\tfrac{315}{8192}\ln 2 - \tfrac{177}{8192}\right)x^{3}$$
$$+\mathrm{O}(x^{4}) + x\ln x\left[\tfrac{3}{16} + \tfrac{9}{512}x + \tfrac{105}{16384}x^{2} + \mathrm{O}(x^{3})\right] . \qquad (7)$$

Note that v($x$) is unusual amongst SMFs in that its exact series expansion requires TWO infinite power series. The superior performance of (6) over the range $0\leq x\leq1$ (see [19]) probably occurs because (6) mimics the form of (7). The form of (7) shows that good approximations cannot easily be obtained by Taylor-expansion or curve-fitting methods, and that two-term approximations, such as (4), cannot provide good general-applicability approximations, although they can perform well over limited ranges of $x$.

### *C. Other field emission special mathematical functions*

Related SMFs can be defined by the equations

$$\mathrm{u}(x) \equiv -\,\mathrm{dv}/\mathrm{d}x, \tag{8a}$$

$$\mathrm{t}(x) \equiv \mathrm{v}(x) + (4/3)\, x\, \mathrm{u}(x), \tag{8b}$$

$$\mathrm{s}(x) \equiv \mathrm{v}(x) + x\, \mathrm{u}(x), \tag{8c}$$

$$\mathrm{r}(x) \equiv \exp[\eta\cdot\mathrm{u}(x)], \tag{8d}$$

where $\eta$ is a "scaling parameter" defined by (10) below. A high-precision (HP) formula exists for $\mathrm{u}(x)$ (see [5]), so high-precision values are easily found for all these functions.

In MG FE theory, when $x$ is replaced by $f$, the last three functions become, respectively, the correction parameter $\mathrm{t}(f)$ in the LECD equation pre-exponential, the slope correction factor $\mathrm{s}(f)$ for a FN plot, and the 2012 intercept correction factor $\mathrm{r}(f)$ [20] for a FN plot.

### *D. Scaled and generalized forms for equations*

The MG FE current-density equation (and some other FE LECD equations) can be written in the "abstract" forms

$$J_{\mathrm{L}} = Z_{\mathrm{F}}\, D_{\mathrm{F}} \approx Z_{\mathrm{F}} \exp[-G_{\mathrm{F}}]\,, \tag{9}$$

where $D_{\mathrm{F}}$ is the tunneling probability for an electron with normal-energy level (i.e. energy perpendicular to the surface) equal to the Fermi level, and $Z_{\mathrm{F}}$ is the related *effective incident current density* (onto the inside of the barrier). When, as in MG FE theory, the simple-JWKB tunneling formalism is used, $D_{\mathrm{F}}$ has the form $\exp[-G_{\mathrm{F}}]$, where $G_{\mathrm{F}}$ is the barrier's *Gamow factor (or " strength")*, as defined below.

By defining $\phi$-dependent "scaling parameters" by

$$\eta \equiv bc^2\phi^{-1/2} \cong 9.836239\, (\mathrm{eV}/\phi)^{1/2}, \tag{10a}$$

$$\theta \equiv ac^{-4}\phi^3 \cong (7.433978{\times}10^{11}\,\mathrm{A\,m^{-2}})\,(\phi\;\;\mathrm{eV})^3, \tag{10b}$$

and by using (5), (2b) can be put in the *scaled form*

$$J_{\mathrm{kL}}{}^{\mathrm{SN}} = \theta f^2 \exp[-\mathrm{v}(f)\cdot\eta/f]\,. \tag{11}$$

A merit is that this form has only one independent variable. It can also be changed into other useful forms (see below).

The "scaled exponent" in (11) can be generalized in the following way. Tunneling barriers can be described by an energy-like parameter $M(z)$, where $z$ denotes distance from the emitter's electrical surface. The barrier strength (or "Gamow factor") $G$ is then defined by

$$G \equiv [2(2m)^{1/2}/\hbar] \int M^{1/2}(z)\,\mathrm{d}z, \tag{12}$$

where $m$ is the mass of the tunneling particle (not necessarily an electron), and the integral is taken "across the barrier", i.e. where $M(z)\geq 0$. A so-called "basic Laurent-form transmission barrier" [5, p. 429] has the mathematical form

$$M(z) = H - Cx - B/x\,, \tag{13}$$

where $H$ and $B$ are constants, and $C$ is a field-related variable. The zeros of $M(z)$, i.e. the "ends" of the barrier, can be written as [5, p.429]

$$z_{\mathrm{ends}} = (H/2C)\,[1 \pm \surd(1-\mu)] \tag{14}$$

where the (dimensionless) *barrier parameter* $\mu$ is defined by

$$\mu \equiv 4BC/H^2. \tag{15}$$

A (dimensionless) generalized scaling parameter $\eta^{\mathrm{GLB}}$ (for a tunneling particle of mass $m$ and a "general Laurent-form barrier") can be defined by

$$\eta^{\mathrm{GLB}} \equiv [(4/3)(2m)^{1/2}/\hbar]\cdot 4BH^{-1/2}\,. \tag{16}$$

It can then be shown, by a lengthy mathematical argument currently spread over several papers, that the barrier strength $G^{\mathrm{GLB}}$ for barrier (13) is given by

$$G^{\mathrm{GLB}} = \mathrm{v}(\mu)\cdot\eta^{\mathrm{GLB}}/\mu\,. \tag{17}$$

Barriers that are special cases of form (13) are used as good or approximate models in several technological contexts, including (obviously) FE from metals, FE from semiconductors, field ionization of gas atoms, post-field-ionization (PFI) of charged metal ions, field evaporation and field desorption. Many nanotechnology techniques use these processes. Equation (17) will allow modern developments in the theory of $\mathrm{v}(x)$ to be applied in these other scientific areas. It will be of particular interest to re-visit the theory of metal-ion PFI, as this has a resemblance to the theory of FE from atomically structured metal surfaces.

### *E. SPME formula for emission current*

Most real emitters have pointed shapes. To deal with this within SPME methodology, one makes the *planar transmission approximation* and (in MG theory) assumes that (2) applies at every surface location "L", with $F_{\mathrm{L}}$ taken as the local surface field. A *characteristic location* "C" is chosen (in modeling, usually at the emitter apex, where the local field is highest if the work function is assumed uniform). Equation (2) is then integrated over the emitter surface to obtain the emission current $I_{\mathrm{e}}{}^{\mathrm{MG0}}$, and the result is written

$$I_{\mathrm{e}}{}^{\mathrm{MG0}} \equiv A_{\mathrm{n}} J_{\mathrm{C}}{}^{\mathrm{MG0}} = A_{\mathrm{n}}\mathrm{t}_{\mathrm{F}}{}^{-2} J_{\mathrm{kC}}{}^{\mathrm{SN}}\,, \tag{18}$$

where the *notional emission area* $A_{\mathrm{n}}$ is defined by (18).

The resulting value of $A_{\mathrm{n}}$ depends on the choice of "C", the LECD equation used, emitter shape and temperature, and the value of the characteristic local field $F_{\mathrm{C}}$ (or characteristic scaled field $f_{\mathrm{C}}$). For emitter-shape effects, there is interest [21] in predicting how the *notional apex-area efficiency* $g_{\mathrm{n}}$ defined by (19) depends on apex scaled field $f_{\mathrm{a}}$:

$$g_{\mathrm{n}}(f_{\mathrm{a}}) \equiv A_{\mathrm{n}}(f_{\mathrm{a}})/2\pi r_{\mathrm{a}}{}^2\,, \tag{19}$$

where $r_{\mathrm{a}}$ is the apex radius of curvature. For example, for a hemisphere on a plane [18,22]

$$g_{\mathrm{n}}(f_{\mathrm{a}}) \approx 1/[\eta/f_{\mathrm{a}} + 4 - \eta/6]\,. \tag{20}$$

### *F. Extended Murphy-Good theory.*

A problem with older SPME methodology is that it disregards the influence that the atomic orbitals of surface atoms must certainly have on the emission process. To recognize this, the present author has introduced a slightly modified form of (2). In (2a), the factor $\mathrm{t}_{\mathrm{F}}{}^{-2}$ is replaced by a *knowledge uncertainty factor* $\lambda_{\mathrm{L}}$ (of unknown functional dependence). A new area-like quantity, the *formal emission area for the SN barrier*, $A_{\mathrm{f}}{}^{\mathrm{SN}}$, is then introduced by

$$A_{\mathrm{f}}{}^{\mathrm{SN}} = \lambda_{\mathrm{C}} A_{\mathrm{n}}\,. \qquad (21$$

Thus, (18) becomes replaced by

$$I_{\mathrm{e}}{}^{\mathrm{EMG}} = A_{\mathrm{f}}{}^{\mathrm{SN}} J_{\mathrm{kC}}{}^{\mathrm{SN}} = A_{\mathrm{f}}{}^{\mathrm{SN}} a\phi^{-1} F_{\mathrm{C}}{}^2 \exp[-\mathrm{v}_{\mathrm{F}} b\phi^{3/2}/F_{\mathrm{C}}]\,. \tag{22}$$

This has been called [23] the *Extended Murphy-Good (EMG) FE equation* (for emission current). The factor $\lambda_{\mathrm{C}}$ takes formally into account all factors neglected in Murphy-Good 1956 FE theory. My current thinking [24] is that $\lambda_{\mathrm{C}}$ is a function of field and most probably lies in the range $0.005<\lambda_{\mathrm{C}}<14$. My view is that vacuum-breakdownVB) simulations should include this uncertainty factor, and experiment with different values for it.

The interpretation of (21) deserves attention. The *notional* emission area $A_n$ relates to the geometrical emission area (which can be defined in various ways), and is the area-parameter theoreticians would normally put into an emission-current prediction or simulation. However, the area-parameter that experimentalists would extract from a FN (or MG) plot (analyzed using a SN barrier) is the *formal* emission area $A_f^{SN}$. These two areas can differ by a large factor. This implies that (at present, and for some years to come) a FE theoretician *cannot* reliably predict what emission current a FE experimentalist will measure, and that a FE experimentalist *cannot* reliably measure geometrical emission area. This implies that FE is a non-typical subject, in that a separation has to be made between "theory for prediction" and "theory for interpretation of experiments". Our interest here is in the latter, which (in the author's view) has historically received much less attention than it deserves.

### F. Murphy-Good-type thermal-field emission theory

Murphy and Good [11] also derived formulae for high-temperature FE, and separate formulae for thermal-field emission. Over the last 20 years, Jensen [18,25,26] has developed a very useful algebraic approximation formula, based on the SN barrier, that spans the whole temperature-and-field range from MG FE theory to classical thermal electron emission theory.

## IV. FE CURRENT-VOLTAGE DATA ANALYSIS

### A. The concepts of "ideal" and "orthodox" systems

Increasing use of non-metallic emitters has led to the realization (probably not yet as widely understood as it needs to be) that measured current-voltage characteristics may often depend on the behavior of the whole electrical circuit and system of which the emitter forms part, not on the emission physics alone (as assumed in most older literature).

A FE device/system is described [5] as *ideal* if measured current-voltage characteristics are determined solely by the system geometry and the emission equations, and not by any other circuit characteristics (e.g., leakage current, series resistance in the current path, or space-charge, amongst other possibilities), *and* if the emitter state (shape and surface work-function characteristics) does not change during system operation, *and* if there are no significant statistical fluctuations in applied voltage or measured current.

Ideal FE devices/systems are further described as *orthodox* if it is adequate to assume that MG FE theory applies, *and* if the relevant local work function is well known. A so-called "orthodoxy test" (see Section IVE) now exists that can indicate whether or not a FE device/system is orthodox. Sections IVB to IVD discuss the analysis of measured current-voltage [$I_m(V_m)$] data taken in orthodox FE situations.

### B. System-geometry-related characterization parameters

By definition, there is no leakage current in an ideal system, so the measured current $I_m$ is equal to the emission current and (for an orthodox system) is assumed given by (6). Hence the *formal* emission area $A_f^{SN}$ is an experimental system characterization parameter.

With emitter arrays, or large-area field electron emitters (LAFEs) generally, the macroscopic (or "LAFE average") emission current density $J_M$ is given by $J_M = I_m/A_M$, where $A_M$ is the LAFE macroscopic area (or "footprint"), and—in orthodox situations—is related to the characteristic kernel current density (for the SN barrier) $J_{kC}^{SN}$ by

$$J_M = I_m/A_M = A_f^{SN} J_{kC}^{SN}/A_M = \alpha_f^{SN} J_{kC}^{SN}, \qquad (23)$$

where $\alpha_f^{SN}$ [$\equiv A_f^{SN}/A_M$] is the *formal area efficiency* (for the SN barrier). Experimental values of $\alpha_f^{SN}$ are not well known, but are thought to typically lie between $10^{-9}$ and $10^{-4}$.

To relate the local field-magnitude $F_C$ to the measured voltage $V_m$, the author prefers to use the formula

$$F_C = V_m/\zeta_C\,, \qquad (24)$$

where $\zeta_C$ is the *voltage conversion length (VCL)*. This relates to the parameter $\beta_V$ sometimes used in FE literature by $\beta_{VC} = 1/\zeta_C$. The VCL is not a physical length but a characterization parameter that assesses how the system geometry affects how easy it is to "turn the emission on". For example, if turn-on occurs at $F_C$ = 3 V/nm (3 GV/m), and the VCL is 1000 nm, then the turn-on voltage ($F_C\zeta_C$) is 3 kV.

The *"macroscopic" (or "applied") field-magnitude* $F_M$ applied to an emitting surface can be related to the measured voltage by $F_M = V_m/d_M$, where the *macroscopic distance* $d_M$ is a calibration parameter which may often be (but does not have to be) a physical distance. A characteristic *field enhancement factor* (FEF) $\gamma_C$ can then be defined by

$$\gamma_C = F_C/F_M = F_C\, d_M/V_m = d_M/\zeta_C\,. \qquad (25)$$

Different system geometries may need slightly different forms of definition for $d_M$, $F_M$ and $\gamma_C$.

### C. The analysis of Fowler-Nordheim-plot slope

Combining (22) and (24), and taking $I_m = I_e^{EMG}$, and then dividing by $V_m^2$ and taking natural logarithms, yields

$$\ln\{I_m/V_m^2\} = \ln\{A_f^{SN} a \phi^{-1} \zeta_C^2\} - \mathrm{v}_F b \phi^{3/2} \zeta_C/V_m\,. \qquad (26)$$

This is the equation for a *theoretical FN plot*, made using natural logarithms. Assuming any voltage dependence in $A_f^{SN}$ or $\phi$ is weak enough to be disregarded, then the slope $S_{FN}(V_m)$ of this theoretical plot (made with natural logs) is

$$S_{FN}(V_m) = \mathrm{d}\ln\{I_m/V_m^2\}/\mathrm{d}(V_m^{-1})$$
$$= -b\phi^{3/2}\zeta_C\, \mathrm{d}(\mathrm{v}_F V_m^{-1})/\mathrm{d}(V_m^{-1})\,. \qquad (27)$$

It follows from the work of Burgess et al. [7] that

$$\mathrm{d}(\mathrm{v}_F V_m^{-1})/\mathrm{d}(V_m^{-1}) = \mathrm{s}(V_m)\,, \qquad (28)$$

where the *slope correction function* (for the SN barrier) was given in [7], as a function of the Nordheim parameter $y_F$, by

$$\mathrm{s}(y_F) = \mathrm{v}(y_F) - \tfrac{1}{2}\, y_F\, \mathrm{dv}/\mathrm{d}y_F\,. \qquad (29a)$$

A "21st-Century" definition, in terms of scaled field $f$, is

$$\mathrm{s}(f) = \mathrm{v}(f) - f\, \mathrm{dv}/\mathrm{d}f\,. \qquad (29b)$$

For an ideal FE system, $f$ is also "scaled voltage", and is related to measured voltage by

$$f = V_m/V_{mR} = V_m/F_R\zeta_C = (c^2\phi^{-2}\,\zeta_C^{-1})\, V_m\,, \qquad (30)$$

where $V_{mR}$ is the *reference measured voltage* (i.e., the $V_m$-value that pulls the barrier top down to the Fermi level. Thus

$$S_{FN}(V_m) = -\,\mathrm{s}(V_m)\cdot b\phi^{3/2}\zeta_C\,, \qquad (31)$$

and the apparent VCL-value $\zeta_{C,app}(V_m)$, at a given measured voltage, can (in principle) be extracted by using the formula:

$$\{\zeta_{C,app}(V_m)\}^{extr} = -\,S_{FN}(V_m)\,/\,[\mathrm{s}(V_m)\cdot b\phi^{3/2}]. \qquad (32)$$

An equivalent FEF-value could then be obtained from (24).

In practical $I_{\mathrm{m}}(V_{\mathrm{m}})$ data analysis, using FN plots, a straight line is fitted to the experimental plot, which may be made using either common or natural logarithms. With natural logarithms, this straight line is written in the form:

$$L^{\mathrm{fit}}(V_{\mathrm{m}}^{-1}) = \ln\{I_{\mathrm{m}}/V_{\mathrm{m}}^{2}\} = \ln\{R_{\mathrm{FN}}^{\mathrm{fit}}\} + S_{\mathrm{FN}}^{\mathrm{fit}}/V_{\mathrm{m}}. \quad (33)$$

With common logarithms, the fitted slope is converted to a value of $S_{\mathrm{FN}}^{\mathrm{fit}}$ by multiplying by ln10, and the parameter $R_{\mathrm{FN}}^{\mathrm{fit}}$ is found by setting the intercept equal to $\log_{10}\{R_{\mathrm{FN}}^{\mathrm{fit}}\}$.

In the *tangent method* of FN plot analysis [12,20], the fitted line is identified with the tangent to the theoretical plot. This tangent has to be taken at a voltage $V_{\mathrm{mt}}$ (and a corresponding, but not initially known, scaled-field value $f_{\mathrm{t}}$) at which the chosen tangent is parallel to the fitted line ($V_{\mathrm{mt}}$ and $f_{\mathrm{t}}$ are called the *fitting values*). Hence a $V_{\mathrm{m}}$-independent "effective VCL" $\zeta_{\mathrm{C,eff}}$ can (in principle) be obtained using

$$\{\zeta_{\mathrm{C,eff}}\}^{\mathrm{extr}} = -S^{\mathrm{fit}} / [\mathrm{s}_{\mathrm{t}} b\phi^{3/2}]. \quad (33)$$

where for notational simplicity $\mathrm{s}_{\mathrm{t}}\equiv\mathrm{s}(f_{\mathrm{t}})$. The function $\mathrm{s}(f)$, given by setting $x$=$f$ in (8c), is a weakly varying function of $f$, and a good first approximation is to take $\mathrm{s}_{\mathrm{t}}^{(1)}$=0.95. Equation (33) then yields a first-approximation VCL-estimate $\zeta_{\mathrm{C,eff}}^{(1)}$, a corresponding FEF-estimate, and [via (30)—provided the choice of $V_{\mathrm{mt}}$ is reasonable], an updated fitting-value estimate $f_{\mathrm{t}}^{(2)}$. Provided that the FE system is ideal, the first-order estimates are often good enough for technology.

If a better result is needed, then—as described in [27]—$f_{\mathrm{t}}^{(2)}$ can be used to generate a new estimate $\mathrm{s}_{\mathrm{t}}^{(2)}$ of slope correction factor, and hence a revised VCL-estimate, $\zeta_{\mathrm{C,eff}}^{(2)}$ and a related revised FEF estimate.

For an ideal FE system, Maxwell's equations predict that the true VCL should be constant. If it is wished to investigate in fine detail whether this is true of a real system (e.g., [28]), then this can be done by using $\zeta_{\mathrm{C,eff}}^{(2)}$ in (30) in order to establish a link between $V_{\mathrm{m}}$ and $f$ (and hence $\mathrm{s}(f)$), and then using (32) with $V_{\mathrm{m}}$-specific values of FN-plot slope. If the extracted VCL is found to depend on $V_{\mathrm{m}}$, then either the FE system is not ideal, or the standard theory given here is incomplete, or that some aspect of it is not exactly correct.

In practice, it might be easier to carry out an investigation of this kind using a Murphy-Good plot [23], and advisable to carry out an orthodoxy test (see below) before doing so.

### *D. Extraction of area values from Fowler-Nordheim plots*

Historically, several methods have been used for extracting area-like parameters from FN plots. The author's currently preferred approach uses the tangent method and an area extraction parameter. This method has been described elsewhere (see Appendix to [23]), so only an outline is given.

In MG FE theory, the theoretical tangent to the theoretical FN plot (22) can be written in the form [20]

$$L^{\mathrm{theor}}(V_{\mathrm{m}}^{-1}) = \ln\{\mathrm{r}(V_{\mathrm{mt}})\cdot A_{\mathrm{f}}^{\mathrm{SN}}\mathrm{a}\phi^{-1}\zeta_{\mathrm{C}}^{2}\} - \mathrm{s}(V_{\mathrm{mt}})\cdot b\phi^{3/2}\zeta_{\mathrm{C}}/V_{\mathrm{m}}, \quad (34)$$

where "r" is a special mathematical function [see (8d)] that acts as the *2012 intercept correction factor for the SN barrier* [20] when $x$ is set equal to $f_{\mathrm{t}}$. In what follows, the abbreviated notation $\mathrm{r}_{\mathrm{t}}\equiv\mathrm{r}(V_{\mathrm{mt}})$ is used. Typically, $\mathrm{r}_{\mathrm{t}}$ has a value in the vicinity of 100.

By identifying the terms in (34) with $\ln\{R^{\mathrm{fit}}\}$ and $S^{\mathrm{fit}}$, it can be shown that the formal emission area (for the SN barrier) $A_{\mathrm{f}}^{\mathrm{SN}}$ can be extracted using the formula

$$\{A_{\mathrm{f}}^{\mathrm{SN}}\}^{\mathrm{extr}} = \Lambda_{\mathrm{FN}}^{\mathrm{SN}}(\phi,f_{\mathrm{t}})\, R_{\mathrm{FN}}^{\mathrm{fit}}\, |S_{\mathrm{FN}}^{\mathrm{fit}}|^{2}, \quad (35a)$$

where the *area extraction parameter* $\Lambda_{\mathrm{FN}}^{\mathrm{SN}}(\phi,f_{\mathrm{t}})$ is given by

$$\Lambda_{\mathrm{FN}}^{\mathrm{SN}}(\phi,f_{\mathrm{t}}) \equiv 1/[(\mathrm{r}_{\mathrm{t}}\mathrm{s}_{\mathrm{t}}^{2})(ab^{2})\phi^{2}]. \quad (35b)$$

In the symbol $\Lambda_{\mathrm{FN}}^{\mathrm{SN}}$, the subscript refers to the form of the plot and the superscript to the tunneling-barrier form assumed when analyzing the data. (The extraction-parameter value depends on what barrier form is assumed.) For a $\phi$= 4.50 eV emitter a typical value of $\Lambda_{\mathrm{FN}}^{\mathrm{SN}}(\phi,f_{\mathrm{t}})$ is 6 nm²/A [29].

The author regards (35) as the simplest method of extracting an emission area from a FN plot relating to an ideal FE system, but this method (and all other older methods) have had weaknesses.

In particular, the derivations of the functions r and s use the function v (see (8)]. Thus, each different approximate algebraic expression for $\mathrm{v}(x)$ leads to a different numerical area estimate [30]. Development of high-precision methods for calculating $\mathrm{s}(x)$ and $\mathrm{r}(x)$ has eliminated this problem.

Further problems (including those related to fitting-value choice) arise because a FN plot is predicted by MG theory to be slightly curved, but experimentalists attempt to fit a straight line. Most of these problems can be diminished by using an alternative plot form—the so-called Murphy-Good (MG) plot [23]—that is predicted (by MG theory) to be "very nearly straight". This method is also described in an ISDEIV 2021 Poster and a separate paper in these Proceedings [31].

### *E. Non-ideal FE systems and the orthodoxy test*

The theory above (and that relating to MG plots) applies to FE systems that are ideal and orthodox. In reality, many FE systems (particularly, but not only, those involving non-metallic emitters) are non-ideal over the whole or part of their operating range, due to one or more "complications". Known complications include: leakage current, series resistance in the current path, current-dependence in FEFs and VCLs, space-charge effects, changes in emitter geometry (e.g., due to Maxwell-stress effects), current-dependent changes in surface state and hence work function, (with non-metals) field penetration into the emitter surface, and possibly (in some contexts) circuit noise.

When a FE system is behaving in a non-ideal manner, then there can be a high probability that extracted characterization parameter values will be spurious. To identify whether or not a FE system is orthodox, Forbes [32] has introduced an *orthodoxy test* that can be applied to measured $[I_{\mathrm{m}}(V_{\mathrm{m}})]$ data, or to converted data using current densities and/or macroscopic fields.

The test is based on the idea that (for an orthodox FE system) a FN or MG plot can be used to measure the characteristic scaled-field value ($f_{\mathrm{C}}$-value) corresponding to any plotted value of the independent variable, denoted here by $X$ and usually a voltage or field, using one of the formulae

$$\{f_{\mathrm{C}}\}^{\mathrm{extr}} = -(\mathrm{s}_{\mathrm{t}}\eta/S_{\mathrm{FN}}^{\mathrm{fit}}) / (X^{-1}), \quad (36a)$$

$$\{f_{\mathrm{C}}\}^{\mathrm{extr}} = -(\eta/S_{\mathrm{MG}}^{\mathrm{fit}}) / (X^{-1}), \quad (36b)$$

where $S^{\mathrm{fit}}$ is the slope of the relevant plot, made against $X^{-1}$.

An analysis [32] of older FE measurements on tungsten emitters (almost always carried out in "ideal" circumstances) has identified typical ranges of $f_{\mathrm{C}}$ in which emitters operate. The complete range $0\leq f_{\mathrm{C}}\leq\infty$ is then divided into "red", "orange" and "green" zones as shown in Fig. 1. (This color-coded notation is due to Popov and the Ioffe Institute group.)

| $f < 0.10$ |
|---|
| $0.10 \leq f < 0.15$ |
| $0.15 \leq f \leq 0.45$ |
| $0.45 < f \leq 0.75$ |
| $0.75 < f$ |

**Fig. 1.** Diagram to illustrate the orthodoxy test, for the $\phi$=4.50 eV case. See text for details. For simplicity, the symbol "$f$", rather than "$f_C$", is used in the diagram.

The range of $f_C$-values that corresponds to the range of $X$-values used in the plot under test is extracted from the FN or MG plot, using (36). An "engineering triage" test is then conducted as follows. (1) If the extracted range falls completely within the "green" zone, then the test is passed. (2) If any part of the extracted range falls within a "red" zone, then the test is failed. (3) In all other cases, the result is inconclusive and more detailed investigation is required.

The ranges shown in Fig. 1 correspond to the $\phi$=4.50 eV case. Slightly different ranges apply to other $\phi$-values [32].

If the orthodoxy test is passed, then the working assumption is that the FE device/system is ideal, and extracted characterization parameters are reliable.

For plots that fail the orthodoxy text, it may be possible to extract an estimated VCL or FEF by "phenomenological adjustment" [33]. Although developed for FN plots, the orthodoxy test should work equally well with MG plots [34].

## V. The Future Development of FE Theory

We now need to return to the situation created by the uncertainty factor in (21), and related comments. Another feature of the situation is that at no stage in the last 90 years have there been reliable, precise quantitative comparisons of theory and experiment relating to FE local current density. Although qualitative trends are predicted correctly, there is a sense in which FE theory is not a truly scientific topic at present, because it has not been justified either by full implementation of the known principles of physics or by precise comparisons with experiment. Further, a former President of the UK Institute of Physics (Prof. Marshall Stoneham, now deceased) thought some of the most difficult unsolved problems in theoretical physics were in FE.

The author's strategy for dealing with this situation involves four main directions of research, as follows.

(1) Consolidate modern SPME methodology, involving:
   (a) active correction of entrenched errors in FE literature;
   (b) encouraging all technological FE areas to move to using modern SPME methodology (or, for simulations, something more sophisticated);
   (c) making small improvements to SPME methodology;
   (d) encouraging consistent terminology and notation use.
   This aims to provide a common starting point for a more scientific approach to field electron emission.

(2) Develop methods of making reliable comparisons between theory and experiment, both in terms of task-oriented theory (e.g., [21]) and experimental procedures.

**(3)** Continue to develop existing FE and TFE theory further, continuing the process of going beyond SPME methodology by incorporating (in particular):
   (a) surface-curvature effects on transmission probability;
   (b) atomic wave-functions and DFT methods;
   (c) lateral coherence and single-atom-emitter effects;
   and by developing
   (d) effective and easily understood algebraic approximate formulae for describing thermal-field emission;
   (e) fast, efficient simulation algorithms;
   (f) the electrostatics of field emitters.
   Obviously, many papers in these areas already exist.

**(4)** Actively explore (or begin to explore) various fundamental and background theoretical problems that are known to exist, but have received little attention.

Sections VI and VII now deal with two issues in wider FE theory, and its application to VB, that may be of interest.

## VI. Pointed Conductors – Feynman Textbook Error

With charged conductors, it has long been known that the electrostatic field is higher in magnitude over sharp or pointed parts of the conductor surface. However, if the extensive literature developed in recent years in the context of field enhancement in field emitter arrays is left aside, then it is difficult to find a simple explanation of this sharp-point effect in general physics literature. The author is aware of slightly complicated discussions presented by Robin [35] and by Starling [36], but these do not lead to simple formulae.

An apparently simple argument is given by Feynman in Vol. II of the textbook based on his Lectures [37]. He models a pointed conductor by using a large sphere (A) of radius $a$ and a small sphere (B) of radius $b$, and argues as follows.

When isolated large and small spheres carry charges of $Q$ and $q$, respectively, the electrostatic potentials at their surfaces (relative to infinity) will be $Q/4\pi\varepsilon_0 a$ and $q/4\pi\varepsilon_0 b$, respectively. If the spheres are well separated but electrically connected (e.g., by a thin wire), the electrostatic potentials will be equal, and hence $q/Q = b/a$. Hence the ratio $F_B/F_A$ of the electrostatic fields at the sphere surfaces (and hence the field enhancement factor $\gamma$) is

$$\gamma \equiv F_B/F_A = (q/4\pi\varepsilon_0 b^2)/(Q/4\pi\varepsilon_0 a^2) = (q/Q)(a^2/b^2) = a/b \ (>1) \,. \quad (37)$$

Feynman argues that the small sphere can represent the tip of a pointed conductor and the large sphere the conductor body, and that (37) shows the field is higher above the "point".

However, we can apply this approach to the well known (in FE) "HCP| model" situation of a cylindrical post with a hemispherical apex, standing on a flat plane of large lateral extent. When $a$ becomes very large, result (37) tends to $\gamma = \infty$, rather than to the correct result (well known in FE, e.g. [38]) that $\gamma \sim h/b,$ where $h$ is the post height. Hence, there must be conceptual inadequacy in Feynman's approach.

An alternative approach applies a method analogous to the "Floating Sphere at Emitter Plane Potential (FSEPP)" model used to discuss field enhancement in FE (e.g., [38,39]). In a low order of approximation, this can be done as follows.

**(1)** On the line joining the sphere centres, select two reference points on the outer sides of the surfaces of the spheres, and let the separation of these reference points be $h$. Consider a configuration with $b \ll h \ll a$.

**(2)** Using only the main terms (due to charges $Q$ and $q$ at the sphere centres), and $k$ to denote $1/4\pi\varepsilon_0$, the condition for the reference points to have the same electrostatic potential is

$$\frac{Qk}{(a+h)} - \frac{Qk}{a} + \frac{qk}{b} - \frac{qk}{(h-b)} = 0 \,. \quad (38)$$

The approximations $a \gg h$, $h \gg b$, yield $q/Q \approx hb/a^2$.

**(3)** The FEF $\gamma$ is then given adequately by

$$\gamma \sim (qk/b^2)/(Qk/a^2) = (q/Q)(a^2/b^2) \approx (hb/a^2)(a^2/b^2) = h/b, \quad (39)$$

which is the usual FSEPP-model approximate formula, sometimes known as the *conducting-post formula*.

Feynman's argument has failed to take into account that the charges on one sphere will influence the electrostatic potential at the surface of the other sphere. His approach would apply if $h \gg a$, but this condition does not provide a good model for a pointed conductor.

Result (39) is better written in the form $\gamma \sim h/r_a$, where $r_a$ is the apex radius of the post, or in the more specific form

$$\gamma = \alpha\, h/r_a\,, \quad (40)$$

where $\alpha$ is an adjustment factor related to the post shape (or, more generally, to the protrusion *and* substrate shapes). Values and functional dependences of $\alpha$ are best evaluated numerically (e.g., [40]). (For the HCP model, $\alpha \sim 0.7$.)

Arguably, the derivation of the conducting-post formula given here is slightly more general than that usually given in FE contexts, and helps to establish that this formula is a *general* approximate result of electrostatics.

The conducting post formula and the general result (40) deserve to be better known in the wider physics community than they currently seem to be, not least because they help to explain the physics of many poorly understood situations involving the Earth's electrostatic field [41]. Amongst others, these situations include lightning rods (where there seems to be limited appreciation that both rod height and apex radius are important), and the electrostatics of trees and mountains.

## VII. ZENER TUNNELING AS A BREAKDOWN MECHANISM

Most treatments of vacuum breakdown in accelerators have assumed that a critical pre-breakdown role is played by FE from clean metallic copper. Others, including the recent BIRD model [42], assume that a surface dielectric layer (most probably an oxide layer) might sometimes be involved.

The theory of FE from bulk semiconductors is reasonably well understood in principle (e.g., [43]), albeit somewhat complicated in detail. If a surface dielectric layer is sufficiently thick, then it seems reasonable to think about vacuum breakdown by using an appropriately adapted version of bulk semiconductor theory.

The most plausible hypothesis is that (as with metals) electron emission is a strictly surface process, involving tunneling through a rounded triangular barrier. The tunneling might take place out of a surface state or out of a triangular well, and the emission current might be limited (mainly or partially) by electron supply rather than by the tunneling barrier. Due to uncertainties about the chemical nature of the dielectric layer, and the actual state of the surface, and because field penetration, band bending and band-structure pinning can occur, it is difficult to know what value to choose for the operating work function. This makes it difficult to estimate what local field value is needed to "turn the emission on", but it seems probable that (as with metal emission) issues will arise about the need to uncover a plausible source of field enhancement, (particularly since a cuprous-oxide-covered surface is reported to have a higher applied breakdown field than clean copper [44]).

An alternative emission mechanism, discussed in semiconductor emission theory [43] as an explanation of observed current-voltage characteristics, but usually considered to take place at much higher local fields than strictly surface emission, is the following.

The primary event, which takes place inside the dielectric, is Zener tunneling [45] by a single electron across the band gap, from the conduction band into valence band. In most cases, this electron is accelerated by the applied field inside the dielectric and creates a swarm of electrons by impact ionization events involving both the primary electrons and secondary electrons, in an "avalanche breakdown" process. If the dielectric layer is of an appropriate thickness, a burst of electrons is emitted into the vacuum, mostly over the top of the transmission barrier. However, some swarm electrons will lose energy inside the dielectric and get trapped (at least temporarily) inside the transmission barrier. These electrons would tend to screen the external field and locally (laterally) diminish the probability of Zener tunneling, at least temporarily. If the excess positive charge inside the dielectric cannot be very quickly neutralized, then the possibility exists of coulomb explosion and local disruption of the dielectric layer. Alternatively, an outcome might be formation of a conducting channel within the dielectric.

There is useful qualitative correspondence between the predicted outcomes of this mechanism and some reports of experimental VB phenomena. Thus, an initial exploration of whether this mechanism is plausible may be useful.

A necessary preliminary is to establish a FE-theory "breakdown onset" criterion. The author's view is that the simplest engineering approach is to first identify a critical value $G^{cr}$ of the barrier strength $G_F^{SN}$ [$=v(f)\cdot \eta/f$], and then attempt to estimate the applied field magnitude at which the barrier strength is reduced to this critical value. This is not an exact approach, but should be good enough to establish whether detailed investigation would be useful.

Conventional thinking [27] is that a tungsten field emitter "turns on" at a characteristic scaled field $f_C \approx 0.20$. A $\phi$=4.50 eV emitter has $\eta \cong 4.637$; $v(f_C=0.20) \cong 0.7440$; $G_F^{SN} \cong 17.25$. This corresponds to a kernel current density (for the SN barrier) of $8.747 \times 10^4$ A/m$^2$. This is equivalent to an tunneling flux density of $5.459 \times 10^{19}$ electrons cm$^{-2}$ s$^{-1}$. However, we are considering a process that in principle can be initiated by single electrons, so maybe a much lower criterion should be chosen, say 100 electrons cm$^{-2}$ s$^{-1}$. This corresponds to $G_F^{SN} \cong 56.1$. It seems useful to explore with the alternative barrier-strength criteria $G^{cr}$ of 17 and 56 in mind.

Strictly, Zener tunneling has its own theory that is different from FE theory. However, it seems an adequate initial approximation to treat the tunneling as that of a particle of effective mass $m_{eff}$ through a SN barrier of height equal to the band gap $E_g$, and of slope $-eF_M/\varepsilon_r$, where $F_M$ is the magnitude of the external applied field, and $\varepsilon_r$ is the dielectric relative permittivity. The resulting $G$-value is

$$G = (m_{eff}/m_e)^{1/2}\, bE_g^{3/2} \varepsilon_r / F_M\,. \quad (41)$$

Thus, the minimum external field necessary to meet the chosen barrier criterion is given (in this simple treatment) by:

$$(F_M)_{min}/(\mathrm{GV/m}) \approx 6.83\,(m_{eff}/m_e)^{1/2}\,\varepsilon_r\,(E_g/\mathrm{eV})^{3/2}/G^{cr}\,. \quad (42)$$

If experimental applied fields are significantly lower, then Zener tunneling is not a credible breakdown mechanism.

Unfortunately, it is difficult to determine reliable values for all the parameters in (42), not least because the precise chemical nature of the dielectric layer is not clearly known. However, one might consider an illustrative set of values [46], for CuO, namely $(m_{eff}/m_e)$=0.58, $\varepsilon_r$ =18.1, $E_g$= 1.35 eV.

For $G^{cr}$=56, this would yield $(F_M)_{min} \approx 2.6$ GV/m. Obviously, this is far higher than the applied fields existing in accelerators, so—on the basis of the above assumed dielectric characteristics for CuO—Zener tunneling does not look plausible as a breakdown mechanism. The main factor in this conclusion is the high assumed relative permittivity.

Obviously, if emission from the dielectric surface built up positive charge at the surface and high internal fields, then possibly this mechanism (or another form of avalanche breakdown) might occur, as suggested by Spada et al. [42].

## VIII. Suggestions for Vacuum Breakdown Science

This paper has outlined a modern version of SPME methodology suited to discussing FE theory and analyzing experiments. It is suggesting that all FE technological areas should move towards using this "21st-Century" methodology as a common basis both for theoretical analysis and for development of a more scientific approach to FE. (But possibly use something more sophisticated for simulations.) In respect of "zero-temperature" FE theory as commonly used, the main changes that I would like to see the Vacuum Breakdown (VB) community implement (where this is not already done) are as follows.

(1) Always use some form of Murphy-Good FE theory, rather than any form of the original 1928/29 FE equation.
(2) State theory using natural rather than common logarithms. [If data are plotted using common logarithms, then convert the slopes and intercepts of fitted lines to "natural log." values].
(3) Use the modern approximations ("F06" and "HP") for "v" and related special mathematical functions.
(4) Move towards the idea of separating the mathematics of "v" from its use in modeling.
(5) Move towards using scaled field $f$, rather than the Nordheim parameter $y$ .
(6) Move towards using Murphy-Good plots rather than FN plots, but use an extraction-parameter approach to extract formal emission areas, whichever plot form is used.
(7) Always apply an orthodoxy test to measured $I_m(V_m)$ data.
(8) Be aware of the need for the "uncertainty factor" $\lambda_C$ in the theory, and include this in simulations.
(9) Distinguish beteween "geometrical", "notional" and "formal" emission areas.
(10) Call the (1956) FE equation used in most VB analysis the *Murphy-Good FE equation*, **not** the "Fowler-Nordheim equation", and do not cite the 1928 FN paper unless you also explain that it contains serious errors.
(11) Do not use the SLAC Report [13] as the only or main reference to FE theory. Also, or instead, cite some modern FE research reference book (e.g., [5] or [18]).

It could be helpful to use the orthodoxy test or (36) to find the $f_C$-value at which emission starts: if significantly above $f_C$=0.20, this could indicate that the emitting feature is **not** a piece of clean metal well connected to the chamber walls.

**E-mail alias of the author: r.forbes@trinity.cantab.net**.